\documentclass[%
 reprint,
superscriptaddress,
 amsmath,amssymb,
 aps,
]{revtex4-2}

\usepackage{graphicx}
\usepackage{dcolumn}
\usepackage{bm}
\usepackage{hyperref}

\usepackage{comment} 
\usepackage{color}

\begin{document}

\preprint{APS/123-QED}

\title{{Operator dependence and robustness of spacetime-localized response in a quantum critical spin chain}
}

\author{Daichi Imagawa}
\affiliation{%
 Department of Physics, College of Humanities and Sciences, Nihon University, Tokyo 156-8550, Japan}
 \author{Keiju Murata}
\affiliation{%
 Department of Physics, College of Humanities and Sciences, Nihon University, Tokyo 156-8550, Japan}
\author{Daisuke Yamamoto}%
\email{yamamoto.daisuke21@nihon-u.ac.jp}
\affiliation{%
 Department of Physics, College of Humanities and Sciences, Nihon University, Tokyo 156-8550, Japan}%
\affiliation{RIKEN Center for Quantum Computing (RQC), Wako, Saitama 351-0198, Japan}

\date{\today}

\begin{abstract}
We investigate the phenomenon of spacetime-localized response in a quantum critical spin system, with particular attention to how it depends on the spatial profile and operator content of the applied perturbation, as well as its robustness against an increase of amplitude and temporal discretization. Motivated by recent theoretical proposals linking such response patterns to the anti–de Sitter/conformal field theory correspondence, we numerically analyze the real-time dynamics of the one-dimensional transverse-field Ising model at criticality using the time-evolving block decimation algorithm. We find that sharply localized and periodically recurring responses emerge only for specific types of perturbations, namely those that correspond to local density fields in the continuum limit. In contrast, perturbations involving other spin components produce conventional propagating excitations without localization. Furthermore, we demonstrate that the response remains qualitatively robust when the time-dependent perturbation is approximated by a piecewise-linear function, highlighting the practical relevance of our findings for quantum simulation platforms with limited temporal resolution. Our results clarify the operator dependence of emergent bulklike dynamics in critical spin chains and offer guidance for probing holographic physics in experimental settings.
\end{abstract}

\maketitle


\section{\label{sec:sec1}Introduction}
Quantum simulation is one of the most promising applications of quantum computing. It allows us to study complex quantum systems that are difficult to simulate with classical computers, mainly because the Hilbert space grows exponentially and quantum entanglement is hard to capture classically~\cite{Feynman1982}. A widely used approach is to employ a controllable quantum system to reproduce the dynamics of another system that is not directly accessible. There are two main types of quantum simulators. Analog quantum simulators, such as those using cold atomic gases~\cite{Bloch2008, Bloch2012,Gross2017-rb} and arrays of Rydberg atoms~\cite{Weimer2010-fu,Barredo2016-ig,Endres2016-yn,Browaeys2020-xe}, are designed to follow the continuous-time evolution governed by a specific Hamiltonian. Digital quantum simulators, such as those using superconducting circuits~\cite{Kim2023-nz,Google_Quantum_AI_and_Collaborators2023-bc,Barends2015} and trapped ions~\cite{Monroe2021-ht,Nguyen2022-ru, Foss-Feig2025-yx}, in contrast, use a programmable sequence of quantum gates to approximate the system's time evolution, with the aim of flexibly simulating a variety of target Hamiltonians. Some physical platforms, including neutral atoms and trapped ions, can be implemented in either analog or digital form depending on the setup. These quantum simulators have made it possible to investigate quantum phase transitions~\cite{Greiner2002-al,Chepiga2021-zg}, nonequilibrium dynamics~\cite{Fukuhara2013-py,Eisert2015-ry,Frey2022-mp,Mi2022-pl}, and various types of strongly correlated behavior~\cite{Islam2013-nm,Leonard2017-ok,Mazurenko2017-xi}.

Beyond condensed matter physics, quantum simulation has been proposed as a powerful tool for exploring a variety of problems in high-energy physics, quantum chemistry, cosmology, and nuclear physics~\cite{Georgescu2014,QS_architectures}. A particularly intriguing direction is the possibility of realizing holographic dualities, such as the anti-de Sitter/conformal field theory (AdS/CFT) correspondence, within controllable quantum systems. This duality relates a gravitational theory in a curved higher-dimensional spacetime to a quantum field theory defined on its lower-dimensional boundary. Although these two sides typically operate at vastly different physical scales, with the AdS bulk corresponding to cosmic regimes and the CFT to microscopic systems, quantum simulators are not restricted by such scale differences. Instead, they allow direct investigation of the structure and dynamics underlying the correspondence by reproducing the essential features of both theories within the same controllable system. This opens the door to experimentally probing aspects of quantum gravity and strongly coupled quantum field theories in tabletop settings, where traditional methods are not available.

In this context, a series of earlier studies involving some of the present authors proposed and examined how geometric features of AdS spacetime can manifest as dynamical phenomena in quantum spin systems~\cite{Kinoshita2023,Bamba2024}. These works were motivated by the observation that the propagation of a massless particle along a null geodesic in AdS spacetime corresponds, in the dual CFT, to a sharply localized signal alternating between distant points~\cite{Kinoshita2023,Terashima2024-wi}. This holographic phenomenon, referred to as the spacetime localized response, was subsequently investigated in a specific spin model, namely the one-dimensional transverse-field Ising model near its critical point~\cite{Bamba2024}. In this setting, a sharply localized signal appears at a distant site after a brief local perturbation and reemerges periodically in time, reflecting the repeated bouncing of a null geodesic between antipodal points on the AdS boundary. The analysis, carried out within the framework of linear response theory using the Jordan–Wigner transformation, clarified that the effect originates from fermionic two-point correlations. 
Although the model corresponds to a CFT with central charge $c = 1/2$ and therefore does not admit a conventional gravitational dual, its continuum limit reduces to a free Majorana fermion theory, which has been suggested to be dual to higher spin gravity~\cite{Ahn:2011pv,Gaberdiel:2011nt}. Thus, while the system lacks a standard gravitational dual, one may still interpret it as probing null geodesics in higher spin gravity through the spin model.

In this paper, we present a more detailed analysis of the spacetime-localized response in the transverse-field Ising model, with a particular focus on how the response depends on the nature of the applied perturbation. Specifically, we compute the real-time dynamics of the transverse-field Ising model on a one-dimensional ring following short-time perturbations, using the time-evolving block decimation (TEBD) algorithm~\cite{Vidal2003-hn,Vidal2004-qo}. We compare the resulting responses for different types of perturbations, including spatially localized perturbations centered at one or multiple sites, as well as perturbations involving different operators. These simulations reveal several key features of the spacetime-localized response. First, the response becomes less sharp than in the linear regime when the perturbation strength is too large, suggesting that the perturbation should remain sufficiently weak in order to observe a clearly localized signal. Second, the response appears only when the perturbation corresponds to the operator associated with particle density in the AdS picture, which strongly reinforces the holographic interpretation. Third, applying perturbations at multiple spatial locations results in independent localized responses, consistent with the expected dynamics of multiple non-interacting particles propagating along null geodesics in AdS spacetime. This research provides a foundation for tabletop quantum simulation experiments aimed at probing aspects of gravitational physics through their correspondence with quantum many-body systems.

The remainder of this paper is organized as follows. In Sec.~\ref{sec:sec2}, we introduce the model, define the spacetime-localized response, and outline its interpretation from the AdS/CFT perspective. Section~\ref{sec:model1} presents our numerical setup and analyzes the response to various spatial perturbations, including single-source, two-source, and spatially uniform excitations. In Sec.~\ref{sec:model2}, we investigate how the response depends on the choice of operator, and explain the differences using field-theoretical correspondences. Section~\ref{sec:discussion} explores the effect of temporal discretization of the source and demonstrates the robustness of the spacetime-localized response under piecewise-linear approximations. Finally, Sec.~\ref{sec:sec6} summarizes our main findings and discusses possible directions for future research.


\section{\label{sec:sec2}Review of Spacetime-localized response}
In this section, we briefly review the concept of the spacetime-localized response~\cite{Bamba2024} and its interpretation based on the AdS/CFT correspondence~\cite{Kinoshita2023,Terashima2024-wi}. The spacetime-localized response refers to a phenomenon in which a sharply localized signal emerges at a distant point after a short-time and spatially localized perturbation is applied to a quantum system. This behavior has been interpreted as a manifestation of holographic duality, where a classical gravitational system in an AdS spacetime is related to a quantum field theory on its boundary.

The gravitational counterpart of this phenomenon is the motion of massless excitations along null geodesics in global AdS spacetimes. For example, in AdS$_{2+1}$ with a unit radius, the spacetime metric is given by
\begin{align}
    ds^2 = -(1 + r^2)\, dt^2 + \frac{dr^2}{1 + r^2} + r^2\, d\phi^2,
    \label{eq:nullgeodesic}
\end{align}
where $t$ is the time coordinate, $r$ is the radial coordinate, and $\phi$ is the angular coordinate along the boundary direction. We set the AdS radius and the speed of light to unity, so all coordinates, including $t$, are dimensionless. The propagation of massless excitations along null geodesics in this geometry can be modeled by a massless scalar field subject to a boundary source $\mathcal{J}(t, \phi)$ with a sharply peaked spatiotemporal profile. A typical choice of the source takes the form~\cite{Kinoshita2023}
\begin{align}
    \mathcal{J}(t,\phi) = A \exp\left[-i\Omega t + iM\phi - \frac{t^2}{2\sigma_t^2} - \frac{\phi^2}{2\sigma_\phi^2} \right],
    \label{eq:source}
\end{align}
where $\Omega$ and $M$ set the central frequency and angular momentum of the injected mode, and $\sigma_t$, $\sigma_\phi$ control its temporal and spatial widths, respectively. {By the above source, a wave packet composed of a bulk field is generated in AdS spacetime. For the eikonal approximation to hold, so that the bulk wave packet can be treated as a null geodesic,  
the bulk field must be localized both in position space and frequency space.  
This condition is given by $1 \ll 1/\sigma_t,1/\sigma_\phi \ll \Omega$. Parameters $\Omega$ and $M$ correspond to the energy and angular momentum of the null geodesic.}
Then, the injected field propagates into the bulk along a null geodesic, traveling from the boundary point $(t, \phi) = (0, 0)$ to the antipodal point $(t, \phi) = (\pi, \pi)$ and bouncing back~\cite{Kinoshita2023}. The points at which a null geodesic collides with the boundary are independent of the angular momentum $M$~\cite{Kinoshita2023,Bamba2024}. The resulting wavepacket continues to oscillate periodically between these boundary points, reflecting the geometry of global AdS.

\begin{figure}[tb]
    \begin{center}
    \includegraphics[width=1.0\linewidth]{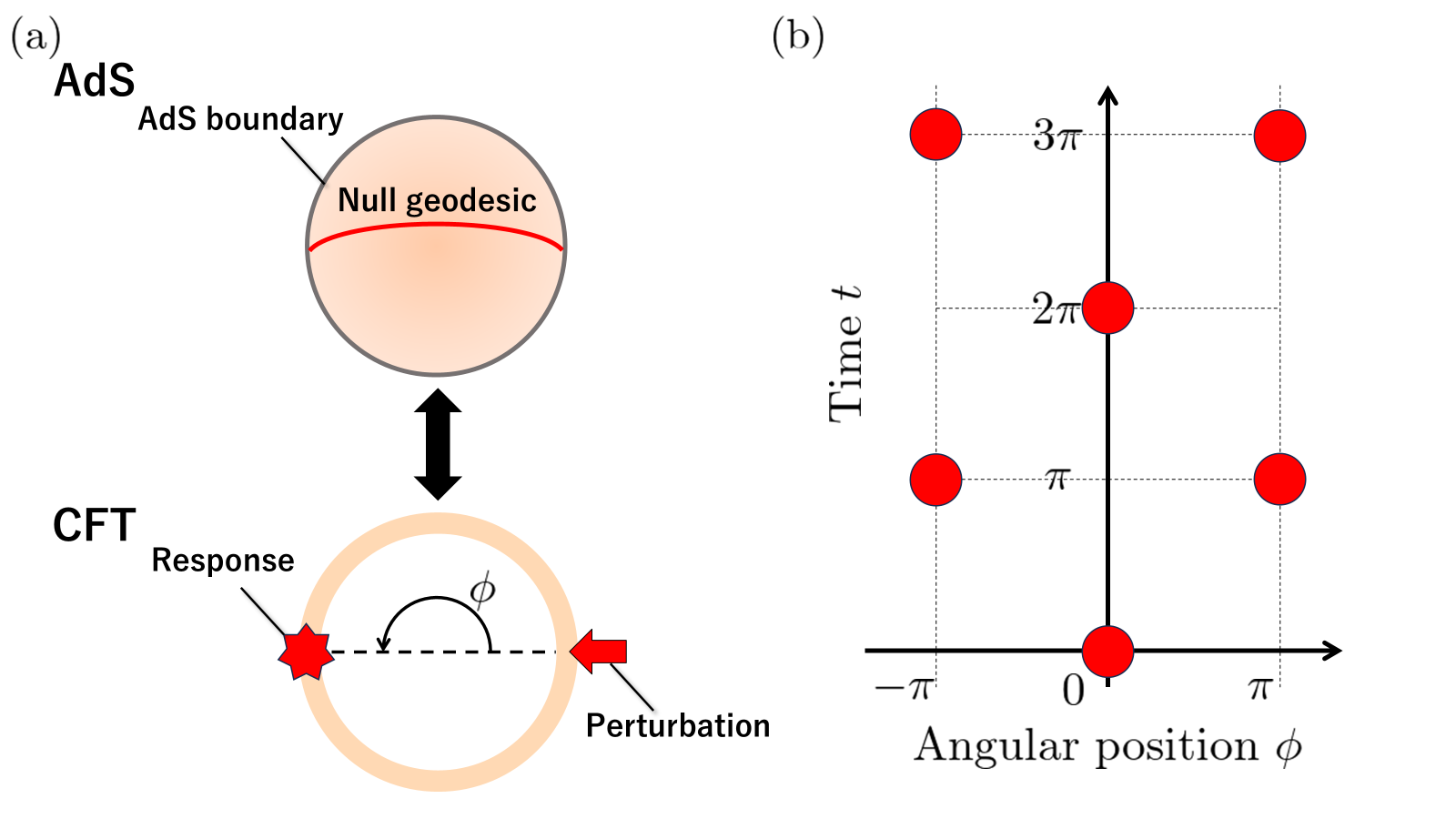}
    \end{center}
    \caption{(a) Schematic illustration of wavepacket propagation along null geodesics in AdS spacetime and its dual manifestation as a spacetime-localized response on the CFT ring, which corresponds to the AdS boundary. (b) Representative spacetime points where the localized response appears. Here $\phi$ denotes the angular position on the CFT ring.}
    \label{fig:example}
\end{figure}

In the framework of the AdS/CFT correspondence, the motion of the wavepacket described above in the bulk should correspond to the emergence of a localized response in the dual CFT, as illustrated in Fig.~\ref{fig:example}(a). When a source $\mathcal{J}(t, \phi)$ is applied at a specific spacetime point on the boundary, the CFT is expected to exhibit a sharply localized signal that appears at a distant location after a finite delay and subsequently reappears periodically~\cite{Kinoshita2023,Bamba2024}, as shown in Fig.~\ref{fig:example}(b). This behavior reflects the causal structure of the bulk spacetime, as encoded in the propagation and reflection of excitations. While the detailed form of the response depends on the operator to which the source couples, the emergence of sharply delayed and spatially separated signals---repeating with a fixed period---is a universal feature in holographic CFTs. This phenomenon is referred to as the spacetime-localized response~\cite{Bamba2024}.

Previous work~\cite{Bamba2024} demonstrated this phenomenon in a concrete setting using a quantum spin chain. Specifically, the transverse-field Ising model on a one-dimensional ring was considered, which is defined by the Hamiltonian
\begin{align}
    H = -J \sum_{j=1}^{L} \sigma_j^z \sigma_{j+1}^z - h \sum_{j=1}^{L} \sigma_j^x,
    \label{eq:Hamiltonian}
\end{align}
where $\sigma_j^{x,z}$ are Pauli matrices acting on site $j$, and periodic boundary conditions $\sigma_{L+1} = \sigma_1$ are imposed. The critical point is realized at $h = J$, where the model is described by a conformal field theory with central charge $c=1/2$. Although this value of $c$ lies outside the usual regime where holographic dualities are well controlled, {in the sense that no standard semiclassical gravity dual is known, the spacetime localized response was still clearly observed~\cite{Bamba2024}. This is because the phenomenon is essentially governed by the causal structure of low order correlation functions, in particular two point functions, which is shared across CFTs with different central charges. In this sense, the classical null geodesic picture in AdS should be understood as a geometric representation of this universal causal structure, rather than as a direct gravitational dual of the spin model.}

To probe the response, a short-time local perturbation was applied to the transverse spin operator $\sigma_j^x$. This choice is motivated by the fact that $\sigma_j^x$ corresponds to the fermion number operator $n_j = c_j^\dagger c_j$ under the Jordan-Wigner transformation, via the relation $n_j = \frac{1}{2}(1 - \sigma_j^x)$ (see Appendix~\ref{appendix:JW}). Accordingly, the perturbation used in Ref.~\cite{Bamba2024} was implemented in the fermionic language as
\begin{align}
    \delta H(t) &= -\sum_{j=1}^{L} \mathcal{J}(t, \phi_j) n_j,
    \label{eq:perturbation_n}
\end{align}
where $\phi_j = \frac{2\pi}{L} \left( j - \frac{L}{2} \right)$ denotes the angular coordinate of site $j$ on the ring. The source function $\mathcal{J}(t, \phi_j)$ is sharply localized both spatially and temporally, with characteristic widths on the order of $\sigma_\phi$ and $\sigma_t$, respectively, ensuring that the perturbation is confined within a small number of sites and a short time interval. This perturbation couples to the operator dual to the bulk scalar field in the holographic picture, and is therefore suitable for realizing the spacetime-localized response.

The change in the expectation value of an operator $O_j$ due to the applied perturbation is defined as $\delta \langle O_j(t) \rangle \equiv \langle O_j(t) \rangle - \langle O_j \rangle_0$, where $\langle \cdot \rangle_0$ denotes the expectation value with respect to the ground state of the unperturbed Hamiltonian. Within linear response theory, the deviation of the fermion number operator $n_j$ in response to the perturbation in Eq.~\eqref{eq:perturbation_n} is given by
\begin{align} 
\delta \langle n_j(t) \rangle = -\sum_{j'}\int_{-\infty}^{\infty} dt' \, G^R_{jj'}(t - t') \, \mathcal{J}(t', \phi_{j'}), 
\label{eq:linearresponse} 
\end{align}
where the retarded Green's function is defined as $G^R_{jj'}(t - t') \equiv -i \theta(t - t') \langle [n_j(t), n_{j'}(t')] \rangle_0$.

This analysis revealed the emergence of a spacetime-localized response in the profile of $|\delta\langle n_j(t)\rangle|$, characterized by sharply delayed and spatially separated signals that periodically reappear due to the causal structure of the dual bulk geometry. These features, schematically illustrated in Figs.~\ref{fig:example}(a) and \ref{fig:example}(b), become increasingly pronounced as the system size $L$ is increased~\cite{Bamba2024}.

In the following sections, we extend this analysis through direct numerical simulations of the transverse-field Ising model. In Sec.~\ref{sec:model1}, we revisit the setup of Ref.~\cite{Bamba2024} beyond the linear response regime and then investigate the effect of introducing multiple perturbations simultaneously. In Sec.~\ref{sec:model2}, we explore how the response behavior changes when different types of operators are used as sources.

\section{\label{sec:model1}{Numerical Analysis: Nonlinear and Multiple Perturbation Effects}}
In this section, we investigate how the spacetime-localized response is modified when going beyond the linear response regime and when multiple local perturbations are introduced simultaneously. {In the context of the AdS/CFT correspondence, incorporating nonlinear effects in the response corresponds to taking into account nonlinear interactions of the bulk fields.  
In the case of the transverse-field Ising model, however, a conventional classical gravitational picture is absent, and it remains unclear whether such an interpretation is valid.  
Here, we simply investigate to what extent nonlinear effects in the response influence the phenomenon of a spacetime-localized response.} We employ the TEBD algorithm to simulate the real-time dynamics of the transverse-field Ising model on a ring. Our analysis includes both single-site perturbations with varying strengths and multi-site perturbations applied simultaneously at different locations. These computations allow us to explore the robustness of the spacetime-localized response under stronger driving and collective excitations.

In the linear response framework, one can formally treat a complex-valued source function $\mathcal{J}(t, \phi)$, which corresponds to a localized wave packet with well-defined frequency and momentum components. However, in real-time numerical simulations of unitary dynamics, the perturbation must be Hermitian, and only real-valued external fields can be directly implemented. Therefore, in our simulations, we consider either the real or imaginary part of the source function separately, as
\begin{align}
    \mathcal{J}_1(t,\phi) &\equiv \mathrm{Re}[\mathcal{J}(t,\phi)] \nonumber \\
    &= A \exp\left[ -\frac{t^2}{2\sigma_t^2} - \frac{\phi^2}{2\sigma_\phi^2} \right] \cos(-\Omega t + M\phi),
    \label{eq:source_1}
\end{align}
\begin{align}
    \mathcal{J}_2(t,\phi) &\equiv \mathrm{Im}[\mathcal{J}(t,\phi)] \nonumber \\
    &= A \exp\left[ -\frac{t^2}{2\sigma_t^2} - \frac{\phi^2}{2\sigma_\phi^2} \right] \sin(-\Omega t + M\phi).
    \label{eq:source_2}
\end{align}
The corresponding real-time dynamics are computed separately under the perturbations
\begin{align}
    \delta H_\eta(t) = -\sum_{j=1}^{L} \mathcal{J}_\eta(t, \phi_j) O_j \quad (\eta = 1,2),
    \label{eq:perturbation}
\end{align}
where the perturbation couples to the operator $O_j$ under consideration. This yields two independent responses of the expectation value $\delta \langle O_j(t) \rangle_1$ and $\delta \langle O_j(t) \rangle_2$. The final signal profile is then constructed as
\begin{align}
    |\delta \langle O_j(t) \rangle| = \sqrt{ \left[ \delta \langle O_j(t) \rangle_1 \right]^2 + \left[ \delta \langle O_j(t) \rangle_2 \right]^2 },
    \label{eq:modulus_general}
\end{align}
which corresponds to the modulus of the complex response to the original source in Eq.~\eqref{eq:source}.

The numerical computation is carried out as follows. Using the unperturbed Hamiltonian defined in Eq.~\eqref{eq:Hamiltonian}, we first obtain the ground state of the system as a matrix product state by applying the density matrix renormalization group (DMRG) algorithm, where the maximum bond dimension is set to $\chi = 600$. 
Starting from this ground state, we simulate the real-time dynamics under the full time-dependent Hamiltonian $H + \delta H_\eta(t)$ using the TEBD algorithm. The initial dimensionless time is chosen as $t = -2$, at which the effect of the source function $\mathcal{J}_\eta(t, \phi_j)$, centered at $t = 0$, is negligible. The time step for the Suzuki–Trotter decomposition is $\tau = 0.01$, and the truncation cutoff in tensor contractions is set to $10^{-11}$. These parameters allow us to accurately track the system’s response to the applied perturbation throughout the entire simulation period. 
The tensor network computations are implemented using the ITensor library~\cite{itensor}.

\begin{figure}
    \includegraphics[width=0.45\textwidth]{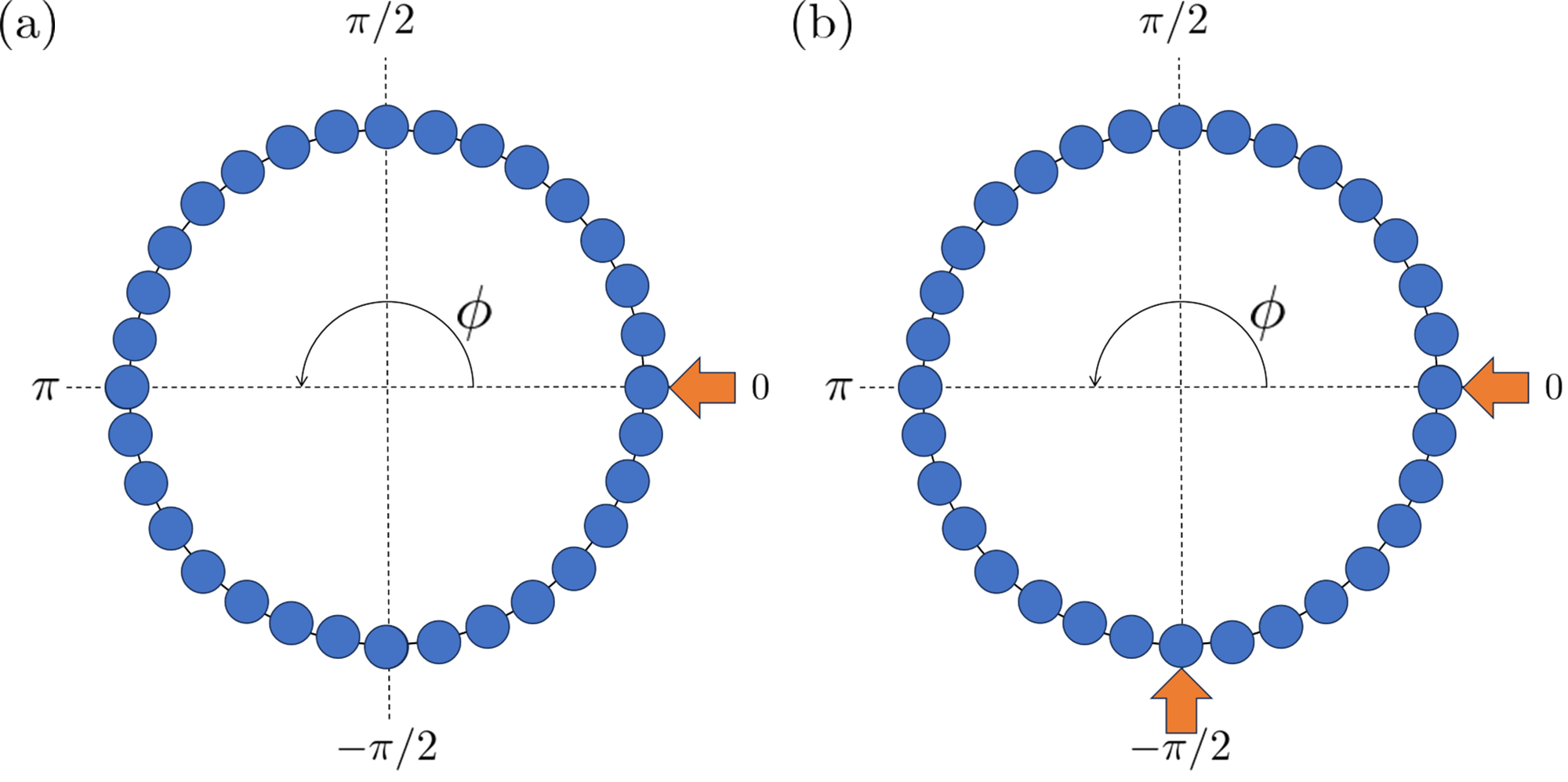}
    \caption{{Central positions of spatially localized perturbations: (a) single source centered at $\phi = 0$; (b) two sources centered at $\phi = 0$ and $\phi = -\pi/2$.}}
    \label{fig:setup_summary}
\end{figure}

In the following subsections, we explore how the spacetime-localized response develops under different spatial profiles of the perturbation. We consider three representative scenarios: (A) a perturbation centered at a single site, (B) simultaneous perturbations applied at two distinct sites, and (C) a spatially uniform perturbation applied across all sites. We assume that the number of sites $L$ is even. In this case, the angular coordinates of sites $j = 1, 2, \dots, L/2, \dots, L$ become $\phi_ j = -\pi + 2\pi/L, -\pi + 4\pi/L, \dots, 0, \dots, \pi$, respectively. A schematic overview of the perturbation setup is provided in Figs.~\ref{fig:setup_summary}(a) and~\ref{fig:setup_summary}(b).

\subsection{Response to perturbations centered on a single site: $\sigma_j^x$ case}

We begin by analyzing the system's response to a spatially localized perturbation centered at a single site. Specifically, we consider the case where the operator $O_j$ in Eq.~\eqref{eq:perturbation} is chosen as the transverse spin operator $\sigma_j^x$, and the spatial profile of the perturbation is centered at site $j = L/2$, i.e., $\phi = 0$ [see Fig.~\ref{fig:setup_summary}(a)]. This setup essentially reproduces the excitation used in Ref.~\cite{Bamba2024}, also described by Eq.~\eqref{eq:perturbation}, with only a factor of $-2$ and a constant shift, neither of which affects the qualitative structure of the response. This allows us to directly examine how the response evolves as the perturbation strength is increased beyond the linear regime.

\begin{figure}
    \includegraphics[width=0.44\textwidth]{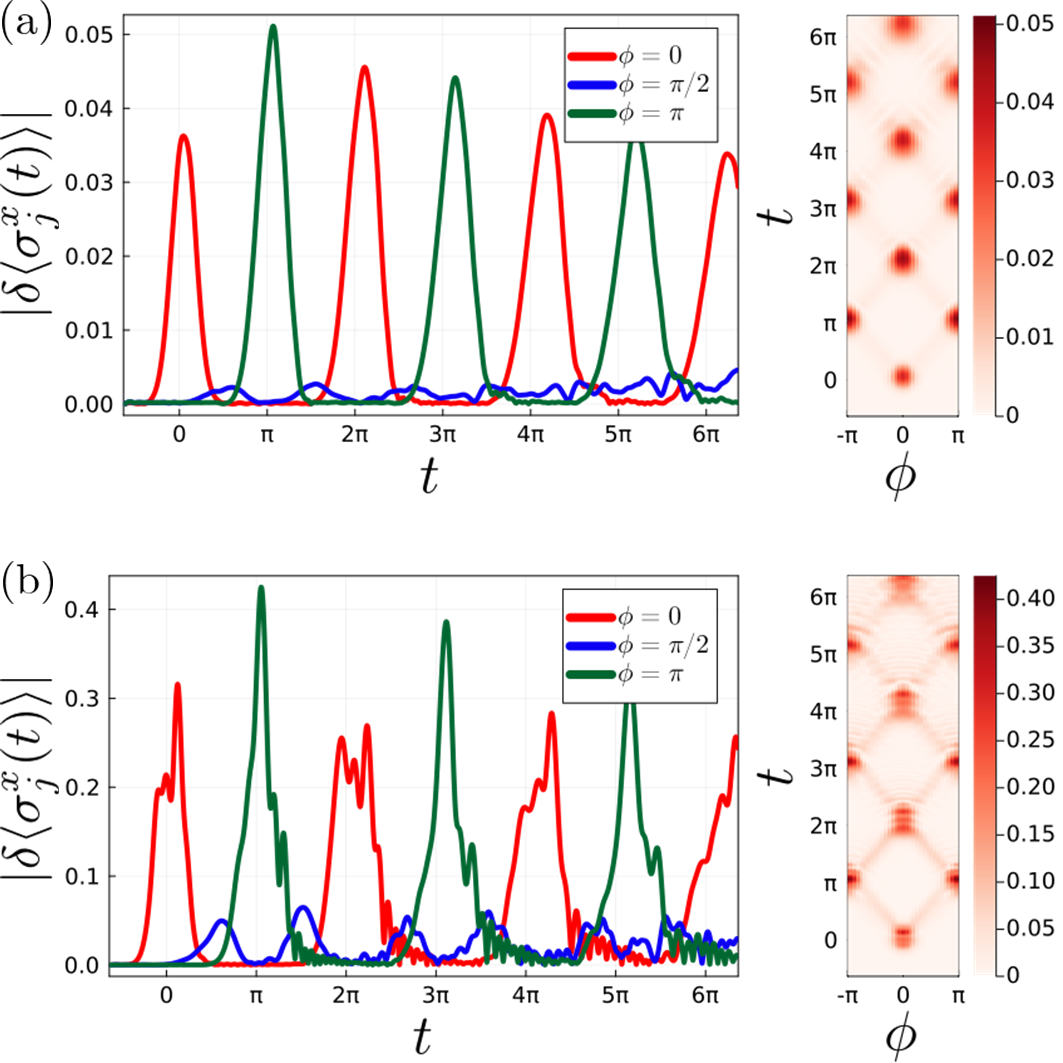}
    \caption{{
Response profiles $|\delta\langle \sigma^x_j(t) \rangle|$ in the transverse-field Ising model on a one-dimensional ring, subject to a perturbation acting on $O_j=\sigma^x_j$, spatially localized and centered at $\phi = 0$. The simulations are performed with parameters $J = h = L / 4\pi$, $\Omega = 5$, $M = 0$, and $\sigma_t = \sigma_\phi = 0.4$, with $L = 32$. (a) Response to a weak perturbation with amplitude $A = 0.1 J \sqrt{2/(\sigma_t \sigma_\phi L)}$. (b) Response to a stronger perturbation with amplitude $A =J \sqrt{2/(\sigma_t \sigma_\phi L)}$. In both (a) and (b), the left panels show the time evolution of $|\delta\langle \sigma^x_j(t) \rangle|$ at three representative angular positions: $\phi = 0$, $\pi/2$, and $\pi$. The right panels display the corresponding spatiotemporal structure of the response $|\delta\langle \sigma^x_j(t) \rangle|$, visualized as intensity plots. 
}}
    \label{fig:fig1}
\end{figure}

We adopt the energy scale convention $J = L / 4\pi$ following Ref.~\cite{Bamba2024}, which naturally ensures that the Fermi velocity becomes $v \equiv 2Ja = 1$, and that the circumference of the one-dimensional ring becomes $\ell \equiv La = 2\pi$ in the continuum limit $L \to \infty$, $a \to 0$, where $a$ is the lattice spacing. The transverse field is fixed at $h = J$ to place the system at the critical point described by a conformal field theory. In this setting, we choose the spatial and temporal widths of the source function to be $\sigma_\phi = 0.4$ and $\sigma_t = 0.4$, respectively, and vary the amplitude $A$. Throughout this paper, we use $L = 32$, which is sufficient to capture the behavior expected from the continuum field theory.

The central frequency $\Omega$ and angular momentum $M$ of the source are set to $\Omega = 5$ and $M = 0$. While these parameters determine the detailed shape of the wave packet in the bulk, they are not essential for the qualitative features of the response on the CFT side corresponding to global AdS~\cite{Bamba2024}.

In the left panel of Fig.~\ref{fig:fig1}(a), we present the time evolution of $|\delta\langle \sigma^x_j(t) \rangle|$ at $\phi = 0$ (the center of the perturbation), $\phi =\pi/2$, and $\phi = \pi$ (the antipodal point) for a perturbation with amplitude $A = 0.1 J\sqrt{2/(\sigma_t \sigma_\phi L)}$, which is sufficiently small to satisfy $\sum_j \int dt\, |\mathcal{J}(t,\phi_j)|^2 \simeq (0.1J)^2$. In this regime, we clearly recover the sharply localized and periodically reappearing response structure predicted by linear response theory: after applying a perturbation centered at $(t, \phi) = (0, 0)$, sharply localized signals appear periodically at $t = 2n\pi$ near $\phi = 0$ and at $t = (2n+1)\pi$ near $\phi = \pi$, where $n$ is a non-negative integer. This is the so-called spacetime-localized response phenomenon, corresponding to null geodesic motion in the bulk AdS geometry, as predicted in previous works~\cite{Kinoshita2023,Bamba2024}. This feature is more clearly visible in the intensity plot of the response amplitude $|\delta\langle \sigma^x_j(t)\rangle|$ shown in the right panel of Fig.~\ref{fig:fig1}(a).

As the amplitude $A$ is increased, however, the response gradually deviates from the ideal spacetime-localized structure predicted by linear response theory. In particular, the signal becomes less sharp and exhibits noticeable spatial broadening. This tendency is clearly observed in Fig.~\ref{fig:fig1}(b), which shows the response amplitude $|\delta\langle \sigma^x_j(t)\rangle|$ for $A = J\sqrt{2/(\sigma_t \sigma_\phi L)}$, ten times larger than the perturbation amplitude used in Fig.~\ref{fig:fig1}(a). While the response still retains peak-like features at the expected times and positions, it develops fine oscillations and irregularities, and the overall localization becomes less distinct. Furthermore, the intensity plot reveals additional linear structures propagating at velocity $v = 1$ along the ring, corresponding to conventional quasiparticle-like excitations. These results indicate that nonlinear effects not only broaden the signal but also give rise to standard propagation modes that obscure the clean causal structure associated with the spacetime-localized response. This highlights the necessity of keeping the perturbation sufficiently weak in order to preserve the characteristic features predicted by the AdS/CFT picture.

\subsection{\label{sec:sigmaX_2}{Response to perturbations centered on two sites: $\sigma_j^x$ case}}

We next investigate the system's response when two spatially localized perturbations are applied simultaneously at distinct positions. {Considering multiple sources is natural in the holographic interpretation, because it corresponds to launching several independent excitations from different boundary points in AdS spacetime. It also provides a test of whether the spacetime-localized response preserves linear superposition in the spin-chain setting. In addition, this two-source configuration serves as an intermediate step toward the case of spatially uniform perturbations discussed in the next subsection, since a uniform drive can be regarded as the limiting case of adding infinitely many localized sources distributed across the entire ring.} Specifically, we consider the case where the operator $O_j = \sigma_j^x$ is perturbed at two sites located at $\phi = 0$ and $\phi = -\pi/2$, illustrated in Fig.~\ref{fig:setup_summary}(b). The perturbation Hamiltonian is given by
\begin{align}
    \delta H_\eta(t) = - \sum_{j=1}^{L} \left[ \mathcal{J}_\eta(t,\phi_j) + \mathcal{J}_\eta(t,\phi_j - \pi/2) \right] \sigma_j^x,
    \label{eq:double_perturbation}
\end{align}
where $\mathcal{J}_\eta(t,\phi)$ is the same Gaussian-modulated envelope used in the single-site case [see Eqs.~\eqref{eq:source_1} and ~\eqref{eq:source_2}]. This setup enables us to explore how the spacetime-localized responses from two separate excitations evolve and interfere, and to assess the extent to which the total response can be interpreted as a linear superposition of individual contributions.

\begin{figure}
    \includegraphics[width=0.45\textwidth]{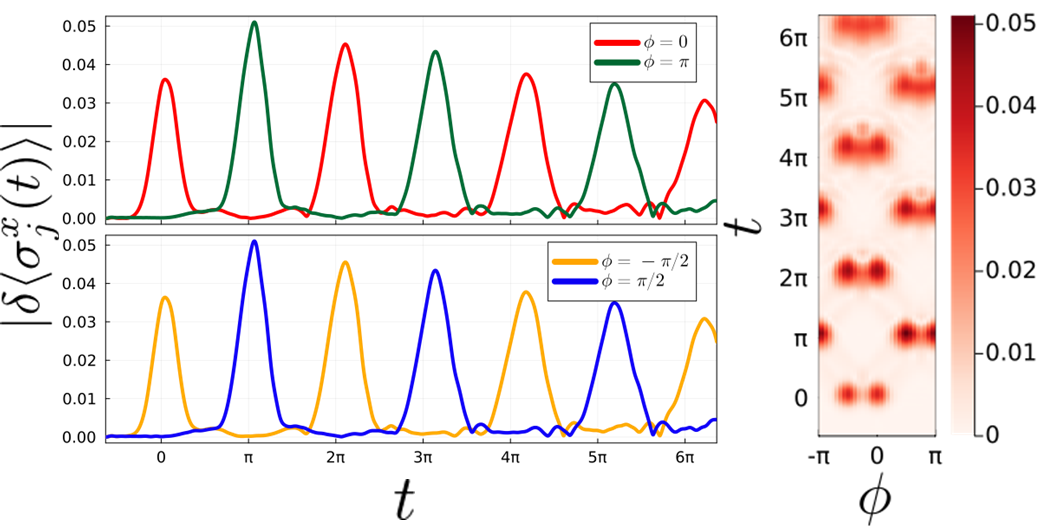}
    \caption{{Response to a weak perturbation acting on $O_j=\sigma^x_j$, spatially localized and centered at two angular positions $\phi = 0$ and $\phi = -\pi/2$, with the same system parameters as those used in Fig.~\ref{fig:fig1}(a). Compared to the single-source case in Fig.~\ref{fig:fig1}(a), multiple localized peaks emerge, corresponding to each excitation and its antipodal counterpart.}}
    \label{fig:fig4}
\end{figure}

Figure~\ref{fig:fig4} shows the resulting response for a weak perturbation with amplitude $A = 0.1 J \sqrt{2/(\sigma_t \sigma_\phi L)}$. Compared to the single-source case, the overall structure remains qualitatively similar, but now exhibits multiple localized peaks reflecting the presence of two independent sources.
Specifically, localized responses appear near $\phi = 0$ and $\phi = -\pi/2$ at even multiples of $\pi$, and near their respective antipodal points $\phi = \pi$ and $\phi = \pi/2$ at odd multiples of $\pi$. This is consistent with the causal structure expected from the AdS side, where two distinct wave packets are launched from different boundary points and follow null geodesics through the bulk. These packets periodically reappear at their antipodal locations and return to their original positions, producing the characteristic spacetime-localized response at multiple positions and times.
This confirms that, in the linear regime, the response remains additive and each localized excitation propagates independently.

\subsection{\label{sec:sigma_X3}{Response to uniform perturbations}}

We now investigate the system's response to a spatially uniform perturbation, in which the same temporal profile is applied equally to all sites. {Studying this case is motivated by two considerations. From the AdS perspective, a uniform drive corresponds to injecting excitations simultaneously from every boundary point, which can be viewed as the limiting situation of adding infinitely many localized sources. From the experimental perspective, uniform perturbations are often easier to implement than spatially resolved ones, and are particularly relevant for platforms where single-site addressability is limited. This configuration can therefore be regarded both as the natural extension of the two-source setup and as a practically important case for quantum simulation.} Specifically, we consider the perturbation operator $O_j = \sigma^x_j$ and apply a uniform source of the form
\begin{align}
    \bar{\mathcal{J}}_\eta(t) \equiv
    \begin{cases}
        A \exp\left(-\dfrac{t^2}{2\sigma_t^2}\right) \cos(\Omega t) & (\eta = 1), \\
        -A \exp\left(-\dfrac{t^2}{2\sigma_t^2}\right) \sin(\Omega t) & (\eta = 2),
    \end{cases}
    \label{eq:uniform_source}
\end{align}
which is independent of the angular coordinate $\phi$. The corresponding perturbation Hamiltonian is given by
\begin{align}
    \delta H_\eta(t) = -\bar{\mathcal{J}}_\eta(t) \sum_{j=1}^{L}  \, \sigma^x_j.
    \label{eq:uniform_perturbation}
\end{align}

\begin{figure}
    \includegraphics[width=0.45\textwidth]{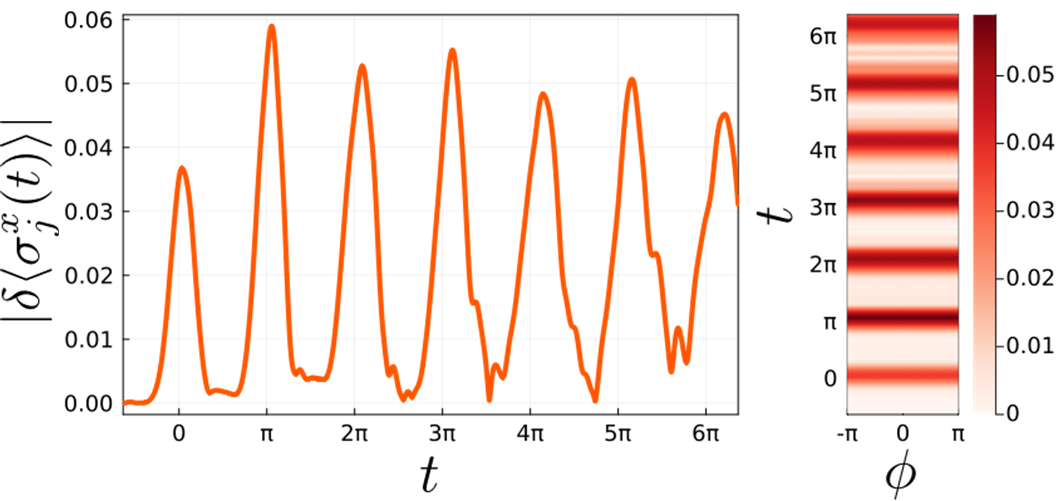}
    \caption{{
    Response to a spatially uniform perturbation acting on $O_j=\sigma^x_j$, corresponding to the limiting case of infinitely many localized sources, with the same system parameters as those used in Fig.~\ref{fig:fig1}(a). {In the left panel, responses from all sites are plotted on top of each other; they coincide within numerical accuracy because the perturbation is uniform.}}}
    \label{fig:uniform}
\end{figure}

The response to this uniform perturbation is shown in Fig.~\ref{fig:uniform}. As expected, the spatial profile of the response is flat across all sites at any given time, while the temporal profile exhibits periodic modulations at intervals of $\pi$. This behavior can be intuitively understood as a simple extension of the two-source case: wave packets are emitted simultaneously from all points on the ring and propagate along null geodesics in the bulk, reaching antipodal points at $t = \pi$, and returning to their original positions at $t = 2\pi$. 

The temporally modulated but spatially homogeneous response arises from the synchronized propagation of independently launched excitations from all sites. This excitation scheme may also be advantageous in experimental platforms where spatially resolved control or measurement is limited, such as quantum simulators lacking single-site addressability.

\section{\label{sec:model2}Dependence of Spacetime-Localized Response on Perturbation Operators}

In the previous section, we focused on perturbations applied to the transverse spin operator $\sigma_j^x$, which is known to exhibit clear spacetime-localized responses consistent with the AdS/CFT correspondence~\cite{Bamba2024}. 
{In earlier work, it was also shown that when the system is tuned slightly away from criticality, the agreement with AdS based expectations becomes weaker, although a remnant of the spacetime-localized response still persists. This raises the question of to what extent the AdS/CFT interpretation remains applicable under different excitation schemes. Here, we explore this issue from a complementary perspective by examining } how the response depends on the choice of perturbation operator.

To this end, we consider localized perturbations involving $\sigma_j^z$ and $\sigma_j^z\sigma_{j+1}^z$. We demonstrate that the emergence of spacetime-localized structures is highly sensitive to the operator type, and explain the physical reasons behind this dependence. Throughout this section, we restrict our analysis to single-source perturbations, as the effects of multiple or extended sources have already been examined in Sec.~\ref{sec:model1}, and their behavior can be inferred from the superposition of individual responses.

\subsection{\label{sec:sigmaZ}Response to $\sigma_j^z$ perturbations}
Let us examine the system's response to a perturbation applied to the longitudinal spin operator $\sigma_j^z$, centered at a single site. The perturbation Hamiltonian is defined by
\begin{align}
    \delta H_\eta(t) = -\sum_{j=1}^{L} \mathcal{J}_\eta(t, \phi_j) \sigma_j^z,
    \label{eq:z_perturbation}
\end{align}
where $\mathcal{J}_\eta(t, \phi)$ is the same Gaussian envelope introduced in Sec.~\ref{sec:model1}, with $\eta = 1, 2$ indicating the real and imaginary parts of the complex source. The response is measured in terms of the same operator, that is, we evaluate $|\delta \langle \sigma^z_j(t) \rangle|$ resulting from the perturbation.

\begin{figure}[tb]
    \includegraphics[width=0.45\textwidth]{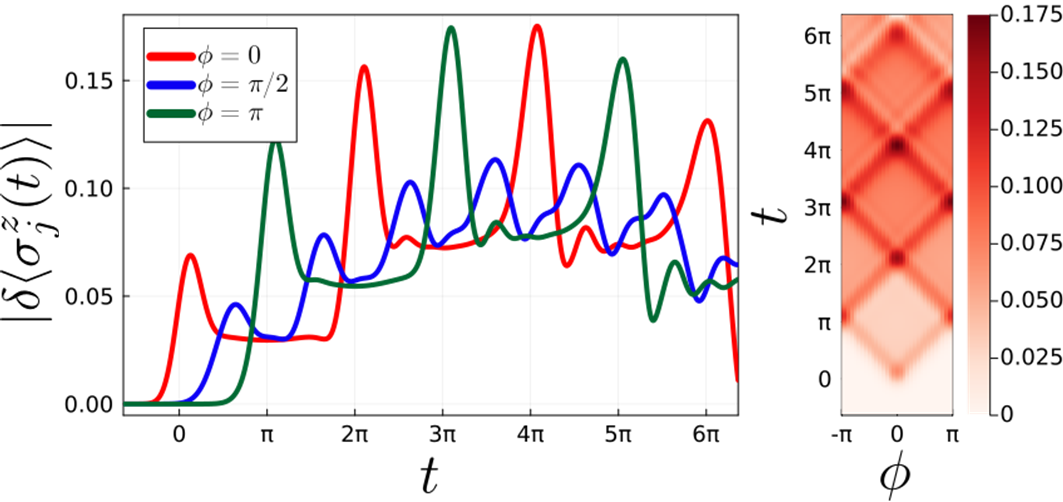}
    \caption{{Response profile $|\delta\langle \sigma^z_j(t) \rangle|$ resulting from a perturbation acting on $O_j=\sigma^z_j$, spatially localized and centered at $\phi = 0$, using the same parameters as in Fig.~\ref{fig:fig1}(a). This figure is directly comparable to Fig.~\ref{fig:fig1}(a), but with the perturbation operator $\sigma_j^x$ replaced by $\sigma_j^z$.}}
    \label{fig:fig2}
\end{figure}

Interestingly, as shown in Fig.~\ref{fig:fig2}, this type of perturbation does not produce a spacetime-localized response. Instead, the signal propagates along the ring at a velocity $v = 1$, consistent with the expected group velocity in the transverse-field Ising model at criticality. This behavior is clearly visible in the intensity plot, which exhibits diagonal streaks in the $(\phi, t)$ plane corresponding to left- and right-moving excitations. Nevertheless, peak structures still appear at integer multiples of $\pi$, both at the origin and at the antipodal point. This can be attributed to the overlap of oppositely propagating wavefronts, which periodically converge at those locations due to the ring geometry.

The stark contrast between the $\sigma_j^x$ and $\sigma_j^z$ cases can be understood from the perspective of operator correspondence in the field-theoretical description. In the transverse-field Ising model, the transverse spin operator $\sigma_j^x$ is related to the fermion number operator via the Jordan-Wigner transformation as $n_j = c_j^\dagger c_j = \frac{1}{2}(1 - \sigma_j^x)$ (see Appendix~\ref{appendix:JW}). In the continuum limit, the fermionic annihilation operator $c_j$ is mapped to the field operator $\Psi(x) = c_j / \sqrt{a}$, where $x = a j$. This correspondence leads to $n_j \sim \Psi^\dagger(x) \Psi(x)/a$, meaning that $\sigma_j^x$ effectively couples to the particle number density in the field-theoretical limit.

In contrast, the longitudinal spin operator $\sigma_j^z$ is mapped, via the Jordan-Wigner transformation, to a nonlocal string operator involving all fermions to the left of site $j$ [see Eq.~\eqref{JWz}]. Because of this nonlocal structure, $\sigma_j^z$ does not correspond to a simple local operator in terms of the fermionic field $c_j$, even before taking the continuum limit. Nevertheless, in the spin representation, $\sigma_j^z$ remains a strictly local operator and produces conventional local excitations that propagate ballistically along the ring. As a result, perturbations in $\sigma_j^z$ excite standard quasiparticle modes rather than inducing the sharply localized, periodically returning signals characteristic of spacetime-localized responses.

This sharp contrast provides further evidence that the emergence of spacetime-localized responses is not a generic property of the spin system, but rather a manifestation of the specific holographic correspondence between certain boundary operators, such as $\sigma_j^x$, and bulk scalar fields in the AdS description, with the caveat that the transverse-field Ising model corresponds to a $c = 1/2$ CFT without a well-defined holographic dual. Still, the correspondence at the level of two-point functions remains meaningful and captures the essential causal structure observed in the response.

\subsection{\label{sec:sigmaZZ}Response to $\sigma^z_j\sigma^z_{j+1}$ perturbations}
We now turn to perturbations involving the nearest-neighbor spin interaction $\sigma_j^z \sigma_{j+1}^z$. While this operator is bilinear in spin variables, it can be shown to correspond to a local density operator in the continuum limit. Using the Jordan-Wigner transformation and expanding in the lattice spacing $a$, we obtain
\begin{align}
    \sigma^z_j \sigma^z_{j+1} &= c^\dagger_j c_{j+1} + c^\dagger_{j+1} c_j + c^\dagger_j c^\dagger_{j+1} + c_{j+1} c_j \nonumber\\
    &= 2a\Psi^\dagger(x)\Psi(x) +\mathcal{O}(a^2),
    \label{eq:ZZ_QFT}
\end{align}
which shows that the leading contribution of $\sigma^z_j \sigma^z_{j+1}$ couples to the fermion number density $\Psi^\dagger(x)\Psi(x)$ as $a \to 0$.

Therefore, much like the $\sigma_j^x$ case, this operator is expected to couple to bulk scalar fields and generate localized wavepacket-like responses in the dual gravitational description, even though it is bilinear in the original spin variables. This case thus serves as a consistency check, reinforcing the operator dependence of the correspondence discussed above.

The perturbation Hamiltonian in this case is defined as
\begin{align}
    \delta H_\eta(t) = -\sum_{j=1}^{L} \mathcal{J}_\eta(t, \phi_j) \, \sigma^z_j \sigma^z_{j+1},
    \label{eq:zz_perturbation}
\end{align}
where $\eta = 1, 2$ indicates the real and imaginary components of the source, as defined in Eqs.~\eqref{eq:source_1} and~\eqref{eq:source_2}. In this subsection, we slightly shift the origin of the angular coordinate and use $\phi_j = \frac{2\pi}{L} \left( j - \frac{L+1}{2} \right)$, so that $\phi = 0$ corresponds to the center of the bond between sites $L/2$ and $L/2+1$, where the perturbation is applied. As in the previous cases, we measure the response in the same operator $\sigma^z_j \sigma^z_{j+1}$. 

\begin{figure}
    \includegraphics[width=0.45\textwidth]{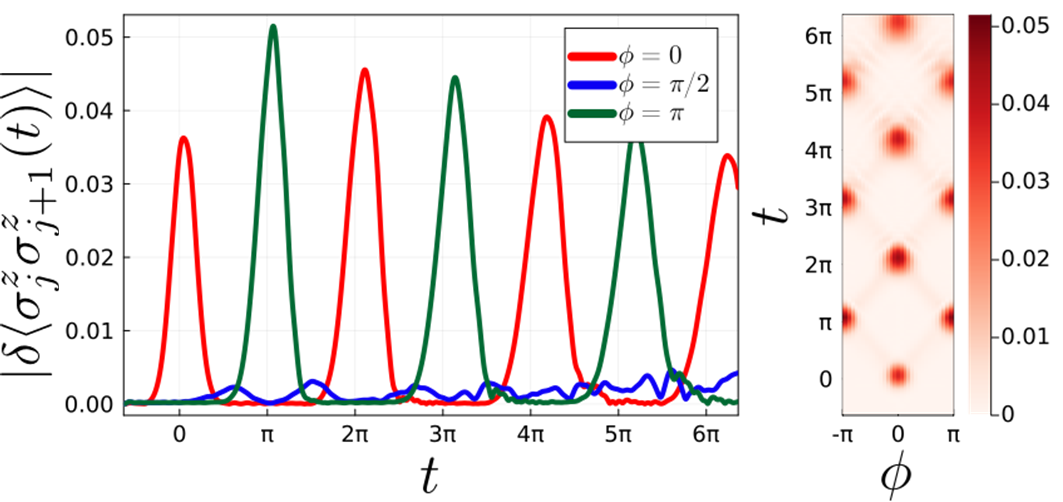}
    \caption{{Response profile $|\delta\langle \sigma^z_j \sigma^z_{j+1}(t) \rangle|$ resulting from a perturbation acting on $O_j=\sigma^z_j \sigma^z_{j+1}$, spatially localized and centered at $\phi = 0$, using the same parameters as in Fig.~\ref{fig:fig1}(a). Sharp, periodically localized responses are observed at $\phi = 0$ and $\phi = \pi$, as in the case of $\sigma_j^x$ perturbations.}}
    \label{fig:fig3}
\end{figure}

As shown in Fig.~\ref{fig:fig3}, this perturbation indeed generates a spacetime-localized response, similar to the case of $\sigma_j^x$. Sharp, periodically reappearing peaks are observed at $\phi = 0$ and $\phi = \pi$ with a period of $\pi$ in time, consistent with the causal propagation of wavepackets along null geodesics in the AdS bulk. This confirms that $\sigma^z_j \sigma^z_{j+1}$ couples to the same type of bulk field as $\sigma_j^x$, despite its bilinear form in the spin representation. 
This result reinforces the key insight that the emergence of spacetime-localized responses is governed not merely by the locality of the spin operator, but by its correspondence to local density-like operators in the continuum field theory. {More generally, the notion of locality relevant for a bulk interpretation should be defined in terms of the effective low energy field theory, rather than the microscopic spin representation. We note that this bulk interpretation becomes meaningful only in the large system size or effectively continuum regime, where lattice scale effects are suppressed, while for small system sizes such effects obscure the emergence of a clear geometric picture.}

It is worth mentioning that the similarity between the $\sigma^x_j$ and $\sigma^z_j \sigma^z_{j+1}$ cases can also be understood in terms of Kramers–Wannier duality~\cite{Kramers, Kramers2, ZAmol}, under which these operators are related. While this duality offers a complementary perspective, the essential reason for the emergence of spacetime-localized responses lies in the fact that both operators couple to local density fields in the continuum limit.
\color{black}

\section{\label{sec:discussion}Discussion}
In the preceding sections, we considered a setting where space is discretized into $L$ sites, while time is treated as a continuous variable. This framework is natural for modeling quantum spin chains and allows high-resolution simulation of real-time dynamics. However, it is both theoretically and practically relevant to examine how the spacetime-localized response is affected when the temporal profile of the perturbation is also discretized. From a theoretical standpoint, such an analysis clarifies how temporal resolution influences the emergence of localized structures. Practically, several quantum hardware platforms, including D-Wave quantum annealers~\cite{Pelofske2020-fx,Pelofske2023-ha} and certain gate-based systems based on superconducting qubits or trapped ions, either restrict the available forms of time-dependent control or favor piecewise-linear implementations for ease of operation. Motivated by these considerations, we have performed additional simulations in which the source function is approximated by a sequence of linear segments in time to assess the robustness of the response.

To isolate the effects of time discretization, we focus on the case $M = 0$, where the source function $\mathcal{J}_\eta(t, \phi)$ factorizes into a product of temporal and spatial components. This separability allows us to approximate the temporal profile independently using a piecewise-linear function, while keeping the spatial Gaussian envelope fixed and centered at $\phi = 0$. Specifically, we replace the oscillatory Gaussian time profiles in Eqs.~\eqref{eq:source_1} and~\eqref{eq:source_2} with piecewise-linear approximations of increasing resolution. For each $\eta = 1, 2$, we construct three variants: one coarse, one intermediate, and one fine. Concretely, we use 2, 4, and 7 segments for the real part ($\eta = 1$), and 3, 5, and 9 segments for the imaginary part ($\eta = 2$), reflecting the symmetry of the respective waveforms.

\begin{figure}[tb]
    \includegraphics[width=0.45\textwidth]{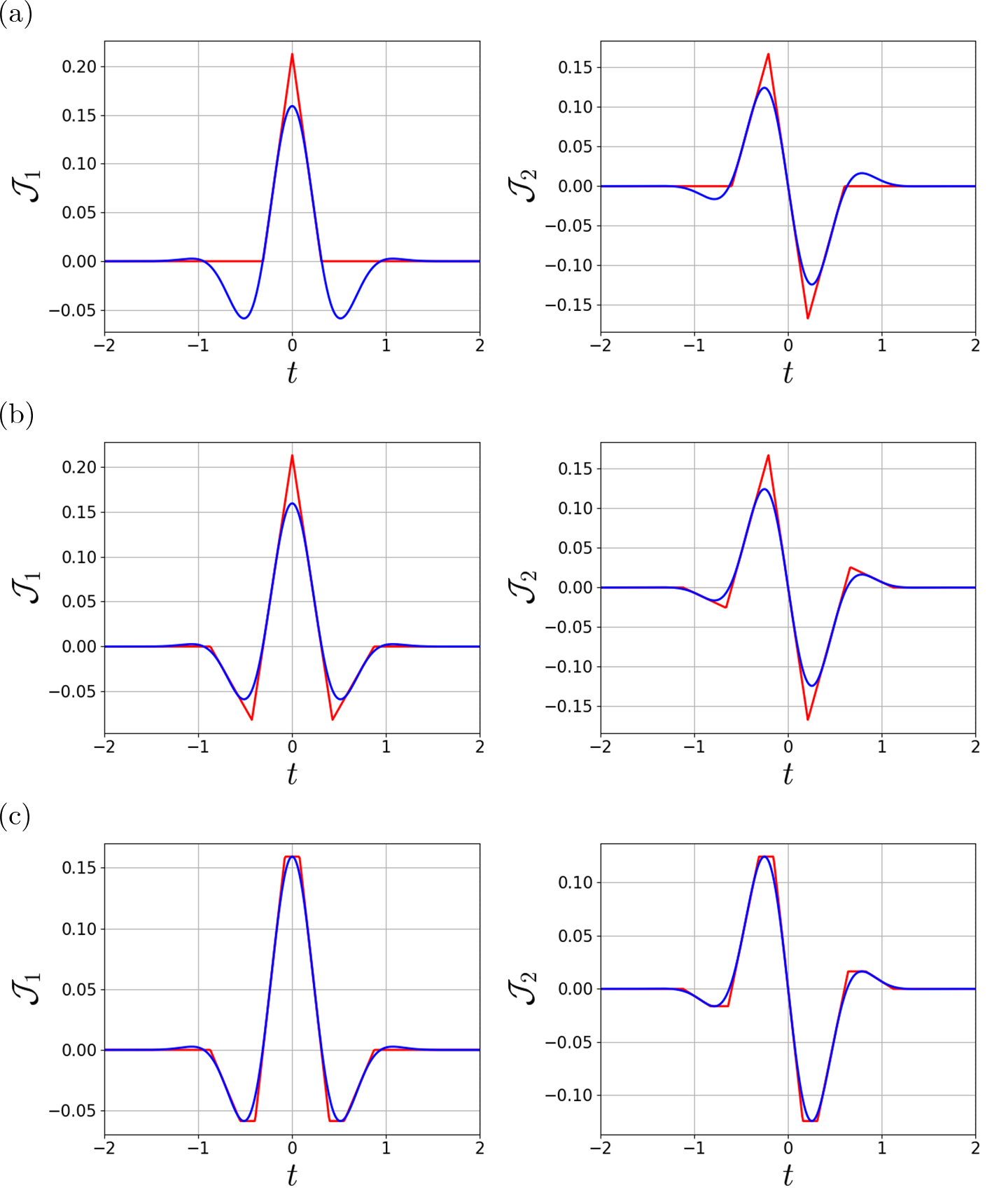}
     \caption{{Piecewise-linear approximations of the real and imaginary parts of the source function $\mathcal{J}_\eta(t,\phi)$ used in the temporally discretized simulations. (a), (b), and (c) present coarse, intermediate, and fine temporal resolutions, corresponding to 2–3, 4–5, and 7–9 segments, respectively. {The blue curves represent the original Gaussian-modulated waveforms, while the red curves represent their piecewise-linear approximations $\tilde{\mathcal{J}}_\eta(t,\phi)$. The plots are shown for $\phi=0$, where the dependence along the $\phi$ direction is simply given by a Gaussian envelope when $M=0$}.} }
     \label{fig:piecewise_source}
  \end{figure}
  
To construct these approximations, we compute the time derivative of the original waveform and identify characteristic points where the derivative vanishes (peaks and troughs) or reaches local extrema in magnitude (regions of steepest slope). Tangent lines are drawn at these points, and the intersections of adjacent tangents are used as vertices for the piecewise-linear function. This procedure captures the essential turning points and rapid variations of the waveform while reducing its complexity. The resulting constructions are illustrated in Figs.~\ref{fig:piecewise_source}(a), ~\ref{fig:piecewise_source}(b), and ~\ref{fig:piecewise_source}(c), respectively.

To evaluate the impact of temporal discretization, we return to the basic setup of Sec.~\ref{sec:model1}, where the perturbation acts on the transverse spin operator $\sigma^x_j$. The perturbation Hamiltonian is modified to incorporate the discretized time dependence as
\begin{align}
    \delta {H}_\eta(t) = -\sum_{j=1}^{L} \tilde{\mathcal{J}}_\eta(t, \phi_j) \, \sigma^x_j,
    \label{eq:piecewise_perturbation}
\end{align}
where $\phi_j = \frac{2\pi}{L} \left( j - \frac{L}{2} \right)$ denotes the angular coordinate of site $j$, and $\tilde{\mathcal{J}}_\eta(t, \phi_j)$ represents the temporally discretized approximation of the original source function. The spatial component is kept unchanged, ensuring that any difference in the resulting response can be attributed solely to the temporal discretization.

\begin{figure}
    \includegraphics[width=0.45\textwidth]{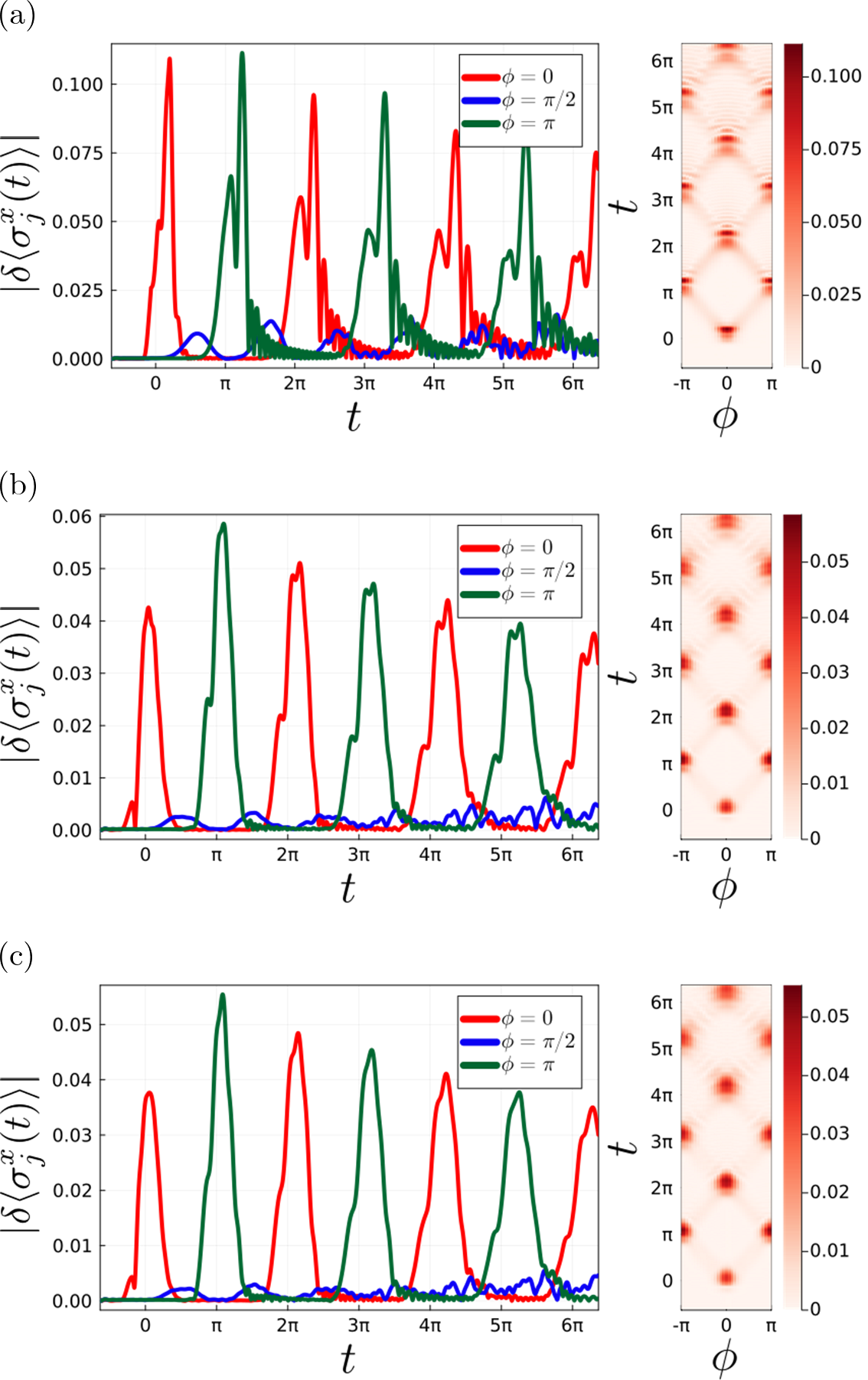}
    \caption{{Spacetime-localized response $|\delta \langle \sigma^x_j(t) \rangle|$ under temporally discretized perturbations, where the source function is spatially localized and centered at $\phi=0$ but its temporal profile is replaced by piecewise-linear approximations. The perturbation acts on $O_j=\sigma^x_j$, with the same system parameters as those used in Fig.~\ref{fig:fig1}(a), except for the temporal profile.
(a) Response with the coarse approximation (2 segments for $\eta = 1$, 3 for $\eta = 2$).
(b) Response with the intermediate approximation (4 segments for $\eta = 1$, 5 for $\eta = 2$).
(c) Response with the fine approximation (7 segments for $\eta = 1$, 9 for $\eta = 2$).}}
    \label{fig:piecewise_response}
\end{figure}

Figures~\ref{fig:piecewise_response}(a)-\ref{fig:piecewise_response}(c) present the resulting dynamics under temporally discretized perturbations with increasing resolution. Panels (a), (b), and (c) correspond to the coarse (2 and 3 segments), intermediate (4 and 5 segments), and fine (7 and 9 segments) piecewise-linear approximations of the real and imaginary parts of the source function, respectively.

Even in the coarsest approximation, the essential spacetime-localized response is clearly visible, with periodically returning signals centered around $\phi = 0$ and $\phi = \pi$. As the temporal resolution improves, the peaks become sharper, and the background noise level diminishes, reflecting closer agreement with the result for the original smooth Gaussian envelope. These findings demonstrate that the spacetime-localized response is remarkably robust against the discretization of the time dependence in the source. In particular, this robustness suggests that even coarse, piecewise-linear implementations may suffice to realize the desired dynamical behavior in experimental settings with limited temporal control.

This observation highlights the practicality of implementing such protocols in hardware platforms where full analog control over time is unavailable or costly, reinforcing the potential utility of our approach in real-world quantum simulation experiments.

\section{\label{sec:sec6}Summary}
In this work, we numerically investigated the emergence of spacetime-localized responses in the transverse-field Ising model at criticality, motivated by recent proposals of an AdS/CFT-inspired correspondence in quantum spin systems~\cite{Bamba2024}. We systematically analyzed how the presence and structure of such responses depend on various aspects of the perturbation, including its spatial profile, amplitude, and operator content.

We began by revisiting single-site and two-site localized perturbations acting on the transverse spin operator $\sigma^x_j$, and confirmed the characteristic signal structures predicted by the correspondence: sharply localized wavepackets periodically reappearing at antipodal points in spacetime, consistent with null geodesic propagation in the AdS bulk. We then extended the analysis to spatially uniform perturbations and found that the localized response is preserved as a coherent collective excitation with periodic returns. 

To probe the operator dependence of this phenomenon, we compared responses induced by perturbations in $\sigma^x_j$, $\sigma^z_j$, and $\sigma^z_j \sigma^z_{j+1}$. Notably, we found that only operators corresponding to local fields in the continuum limit, such as $\sigma^x_j$ and $\sigma^z_j \sigma^z_{j+1}$, give rise to spacetime-localized responses. In contrast, perturbations in $\sigma^z_j$ generate conventional propagating signals that spread along the ring, reflecting the nonlocal fermionic structure of the operator. This distinction reinforces the idea that spacetime-localized responses emerge only when the boundary operator couples to a local scalar field in the effective bulk theory.

In the final part of this study, we investigated the robustness of the spacetime-localized response under temporal discretization of the perturbation. Motivated by the control constraints in some experimental platforms, we replaced the smooth time profile of the perturbation with piecewise-linear approximations of varying resolution. We found that even coarse approximations preserve the key features of the response, suggesting that high temporal fidelity is not strictly required for observing this phenomenon. This result may facilitate future experimental realization on noisy or constrained hardware.

While our study provides strong support for the AdS/CFT-inspired interpretation, we emphasize that the transverse-field Ising model has central charge $c = 1/2$, and thus its holographic interpretation, at least in terms of semiclassical gravitational duals, is expected to be limited to low-order correlators. The emergence of spacetime-localized responses in this case should be regarded as a highly nontrivial consistency check, rather than evidence of a full-fledged dual geometry. As a direction for future work, it would be of interest to explore systems with larger central charge, such as SU($N$) Heisenberg chains~\cite{Nataf2018-ag} or other multicomponent models, where richer holographic structure might become visible, potentially including phenomena such as finite-temperature black hole physics~\cite{Hashimoto:2019jmw,Hashimoto:2018okj,Kaku:2021xqp,Hashimoto:2022aso}.

\begin{acknowledgments}
{We would like to thank H.~Akishima, K, Endo, T. Kadowaki, Y. Suzuki, and S. Kinoshita for useful discussions. The work of D.Y. was supported by JSPS KAKENHI Grants No.~21H05185, No.~23K25830, No.~24K06890, and JST PRESTO Grant No.~JPMJPR245D.} 
\end{acknowledgments}

\appendix
\vspace{2ex}
\section{Jordan-Wigner transformation}
\label{appendix:JW}

The transverse-field Ising model can be mapped onto a system of spinless fermions via the Jordan-Wigner transformation. In this mapping, the Pauli operators are expressed in terms of fermionic creation and annihilation operators as
\begin{align}
    \sigma_j^x &= 1 - 2 c_j^\dagger c_j, \label{JWx}\\
    \sigma_j^z &= \left( \prod_{k<j} (1 - 2 c_k^\dagger c_k) \right) (c_j + c_j^\dagger),\label{JWz}
\end{align}
where $c_j$ and $c_j^\dagger$ are fermionic annihilation and creation operators at site $j$, satisfying the anticommutation relations $\{ c_j, c_k^\dagger \} = \delta_{jk}$ and $\{ c_j, c_k \} = \{ c_j^\dagger, c_k^\dagger \} = 0$.

The nonlocal string operator in $\sigma^z_j \sigma^z_{j+1}$ cancels out for $j < L$, and the interaction term reduces to
\begin{align}
    \sigma^z_j \sigma^z_{j+1} = c^\dagger_j c_{j+1} + c^\dagger_{j+1} c_j + c^\dagger_j c^\dagger_{j+1} + c_{j+1} c_j.
    \label{eq:JW_zz}
\end{align}
This expression includes nearest-neighbor hopping and pairing terms. Using this result, the transverse-field Ising model can be rewritten as
\begin{align}
    H &= -J \sum_{j=1}^{L} (c_j^\dagger c_{j+1} + c_{j+1}^\dagger c_j + c_j^\dagger c_{j+1}^\dagger + c_{j+1} c_j) \nonumber \\&- h \sum_{j=1}^{L} (1 - 2 c_j^\dagger c_j).
\end{align}

When periodic boundary conditions are imposed on the spin system, the corresponding boundary term $\sigma^z_L \sigma^z_1$ in the fermionic representation involves a Jordan-Wigner string that depends on the total fermion number parity. As a result, the fermionic boundary condition becomes parity-dependent: it is periodic in the even-parity sector and anti-periodic in the odd-parity sector. To maintain a local Hamiltonian form, it is common to fix the parity sector in practical calculations.


\providecommand{\noopsort}[1]{}\providecommand{\singleletter}[1]{#1}%

\end{document}